\documentclass[conference]{IEEEtran}
\IEEEoverridecommandlockouts
\usepackage{cite}
\usepackage{amsmath,amssymb,amsfonts}
\usepackage{algorithmic}
\usepackage{graphicx}
\usepackage{textcomp}
\usepackage{xcolor}
\usepackage{siunitx}
\usepackage{url}
\usepackage{cleveref}
\usepackage{subfig}
\usepackage{comment}

\Crefname{figure}{Fig.}{Figs.}
\Crefname{equation}{Equation}{Equation}
\crefname{equation}{}{}

\newcommand{\sinc}[0]{\text{sinc}}
\newcommand{\fenc}{f_\text{enc}}
\newcommand{\fdec}{f_\text{dec}}
\newcommand{\fmask}{f_\text{mask}}
\newcommand{\fnomask}{f_\text{no\_mask}}
\newcommand{\fnn}{f_\text{NN}}

\newcommand{\jdec}{J_\text{dec}}
\newcommand{\jmask}{J_\text{mask}}
\newcommand{\Fs}{F_\text{s}}

\def\BibTeX{{\rm B\kern-.05em{\sc i\kern-.025em b}\kern-.08em
    T\kern-.1667em\lower.7ex\hbox{E}\kern-.125emX}}
\begin{document}

\title{Local Equivariance Error-Based Metrics for Evaluating Sampling-Frequency-Independent Property of Neural Network
\thanks{This work was supported by JSPS KAKENHI under Grant JP23K28108 and JST ACT-X under Grant JPMJAX210G.}
}

\author{\IEEEauthorblockN{Kanami Imamura$^{1,2}$,  Tomohiko Nakamura$^2$, Norihiro Takamune$^1$, Kohei Yatabe$^3$, and Hiroshi Saruwatari$^1$}
\IEEEauthorblockA{$^1$ Graduate School of Information Science and Technology, The University of Tokyo, Tokyo, Japan}
\IEEEauthorblockA{$^2$ The National Institute of Advanced Industrial Science and Technology (AIST), Japan}
\IEEEauthorblockA{$^3$ Department of Electrical Engineering and Computer Science, \\ Tokyo University of Agriculture and Technology, Tokyo, Japan}
}

\maketitle

\begin{abstract}
Audio signal processing methods based on deep neural networks (DNNs) are typically trained only at a single sampling frequency (SF) and therefore require signal resampling to handle untrained SFs.
However, recent studies have shown that signal resampling can degrade performance with untrained SFs.
This problem has been overlooked because most studies evaluate only the performance at trained SFs.
In this paper, to assess the robustness of DNNs to SF changes, which we refer to as the SF-independent (SFI) property, we propose three metrics to quantify the SFI property on the basis of local equivariance error (LEE).
LEE measures the robustness of DNNs to input transformations.
By using signal resampling as input transformation, we extend LEE to measure the robustness of audio source separation methods to signal resampling.
The proposed metrics are constructed to quantify the SFI property in specific network components responsible for predicting time-frequency masks.
Experiments on music source separation demonstrated a strong correlation between the proposed metrics and performance degradation at untrained SFs.
\end{abstract}


\section{Introduction}
Deep neural networks (DNNs) have been widely used for various audio signal processing tasks~\cite{Hendrik2019JSTSP}.
In these tasks, DNNs are typically trained and evaluated only at a single sampling frequency (SF).
To apply the trained models to signals with different SFs, signal resampling is usually required.
Although resampling is well-grounded in signal processing theory~\cite{Eldar2015samplingtheory}, recent studies have shown that it can degrade performance at untrained SFs in DNN-based audio source separation methods~\cite{saito2022sficonvtasnet,imamura2023noninteger,yu2023sfibsrnn,zhang2023uses,zhang2024urgent}.
This observation underscores the need to assess not only separation performance at trained SFs but also how robust these methods are to SF changes---that is, how independent they are of SFs.
We refer to this independence as \emph{the SFI property} and address its evaluation in this paper.

Although there are well-established evaluation metrics for separation performance~\cite{vincent2006sdr,roux2019sisdr}, there is currently no standard metric for evaluating the SFI property.
For separation performance, source-to-distortion ratio (SDR)~\cite{vincent2006sdr} and scale-invariant SDR (SI-SDR)~\cite{roux2019sisdr} are commonly utilized in research on audio source separation literature~\cite{araki2025ass}.
Previous studies have measured the SFI property by calculating these metrics at various SFs~\cite{saito2022sficonvtasnet,paulus2022sfi,yu2023sfibsrnn,zhang2023uses,imamura2023noninteger,zhang2024urgent,imamura2024naf}.
However, while this evaluation is useful for comparing the performance of audio source separation methods at each test SF, it does not quantify \emph{how} SFI a method is.
Since SDR and SI-SDR depend on SFs by definition, their values at different SFs cannot be directly compared.
Therefore, we need to construct an evaluation metric that quantifies the SFI property and explains the performance degradation at untrained SFs.

In this paper, we propose evaluation metrics to quantify the SFI property by measuring the robustness of audio source separation methods to signal resampling.
The proposed metrics are based on local equivariance error (LEE)~\cite{gruver2023lee}.
Originally introduced for image processing tasks, LEE measures the robustness of DNNs to input transformations such as rotation and translation.
Since signal resampling is also a form of input transformation, we extend LEE to quantify the SFI property.
Through this extension, we find that directly applying LEE does not explain the performance degradation at untrained SFs, as discussed in \Cref{sec:result-sdr}.
To resolve this issue, the proposed evaluation metrics target not the entire network but specifically the network components responsible for predicting time-frequency masks.
Through experiments on music source separation, we will analyze the relationship between the proposed metrics and performance degradation at untrained SFs.

\section{Related Works}
\subsection{Local Equivariance Error} \label{sec:lee}
LEE is defined on the basis of \emph{equivariance}, a property of a function where a transformation applied to the input space results in an equivalent transformation in the output space.
Equivariance in DNNs has proven effective for tasks such as image line drawing and semantic segmentation~\cite{chen2023equivariance}.

Let $f:\mathbb{V}\rightarrow \mathbb{V'}$ be a function and $x\in\mathbb{V}$ be its input sampled from a data distribution $\mathcal{D}$, where $\mathbb{V}$ and $\mathbb{V'}$ are input and output vector spaces, respectively.
We say that $f$ is equivariant under the action of a group $\mathcal{G}$ if the following equation holds for any $\mathsf{g} \in \mathcal{G}$:
\begin{equation}
    f(\tau(\mathsf{g})x) = \tau'(\mathsf{g})(f(x)),
    \label{eq:equivariance}
\end{equation}
where $\tau(\mathsf{g}):\mathbb{V}\rightarrow\mathbb{V}$ and $\tau'(\mathsf{g}):\mathbb{V'}\rightarrow\mathbb{V'}$ are the representations of $\mathsf{g}$ for the input and output of $f$, respectively.
For example, we consider the rotation group $\mathcal{G}$ for images and a group element $\mathsf{g}_r$ smoothly connected on $\mathcal{G}$ by the parameterization of angle $r$.
The realizations of $\mathsf{g}_r$ in the input and output domains ($\tau(\mathsf{g}_r)$ and $\tau'(\mathsf{g}_r)$) are rotation matrices corresponding to angle $r$.
When $r=0$, they reduce to identity matrices, i.e., no rotation.
For simplicity, we denote $\tau(\mathsf{g}_r)$ and $\tau'(\mathsf{g}_r)$ as $g_r$ and $g'_r$.

LEE measures the degree of equivariance in a DNN.
We assume that $g'_r$ is invertible and define $h_r(x)=g'^{-1}_r(f(g_r(x)))$.
The difference between $h_r(x)$ and $f(x)$ serves as a measure of deviation from equivariance.
Taking the limit as $r\to0$, LEE is given as
\begin{align}
\text{LEE}(f) &= \dfrac{1}{V}\,\mathbb{E} [|| \mathcal{L} f(x) ||^2], \label{eq:lee} \\
\mathcal{L}f &= \lim_{r\to0} \dfrac{h_r - f}{r} = \left. \dfrac{d}{d r} h_r  \right|_{r=0},\label{eq:derivative} 
\end{align}
where $\mathbb{E}$ denotes the expectation with respect to $x$ over the distribution $\mathcal{D}$ and $V$ is the output dimension of $f(x)$.
$\mathcal{L}f$ is called the Lie derivative of $f$, which measures the degree of change in $f$ under an infinitesimal transformation.

LEE was extended to approximately compute the contributions of specific layers using the chain rule of the Lie derivative~\cite{gruver2023lee}.
We refer to this extension as \emph{layerwise LEE}.
Let $\fnn(x)=(f_I \circ f_{I-1} \circ \ldots \circ f_1)(x)$ be a DNN composed of $I$ layers, where $f_{i}$ $(i=1,\ldots,I)$ denotes the function representing layer $i$.
By applying the chain rule, the Lie derivative of $\fnn$ can be decomposed as
\begin{equation}
    \mathcal{L}\fnn = \sum_{i=1}^I J_{I:i+1} \mathcal{L}f_i, \label{eq:layerwise_lie}
\end{equation}
where $J_{I:i+1}$ is the Jacobian matrix of $f_I \circ \ldots \circ f_{i+1}$ and $J_{I:I+1}$ is an identity matrix.
For notational simplicity, we omit the input of each Lie derivative and Jacobian matrix.
From \cref{eq:layerwise_lie}, we derive the upper bound of the norm of $\mathcal{L}\fnn$:
\begin{equation}
    ||\mathcal{L}\fnn|| \leq \sum_{i=1}^I || J_{I:i+1} \mathcal{L}f_i ||. \label{eq:layerwise_lee}
\end{equation}
Each term on the right-hand side can be interpreted as the contribution of an individual layer.
The layerwise LEE of layer $i$ is defined as the expectation of the term associated with $f_i$ in \cref{eq:layerwise_lee} and serves as a tool for analyzing the influence of each layer on LEE.
See \cite{gruver2023lee} for further details on LEE and layerwise LEE.

\subsection{TasNet Architecture} \label{sec:tasnet}
The TasNet architecture~\cite{luo2018tasnet} is one of the foundational network architectures for DNN-based audio source separation and is widely utilized in literature~\cite{luo2019convtasnet,saito2022sficonvtasnet,luo2020dprnn}.
It consists of an encoder $\fenc$, a decoder $\fdec$, and a mask predictor $\fmask$.
The encoder and decoder serve as a trainable time-frequency transform and its inverse transform, respectively.
The encoder transforms the input signal into a pseudo time-frequency representation, which is passed to the mask predictor.
The mask predictor predicts masks of individual sources and multiplies the representation by these masks.
The masked representations are then fed into the decoder to produce the separated signals.
In this paper, we explore the DNN with the TasNet architecture and denote it as $\fnn=\fdec \circ \fmask \circ \fenc$.

\section{Proposed Metrics} \label{sec:proposed-lee}
\subsection{LEE Using Resampling as Input Transformation}
\begin{figure}[t]
    \centering
    \includegraphics[width=\linewidth]{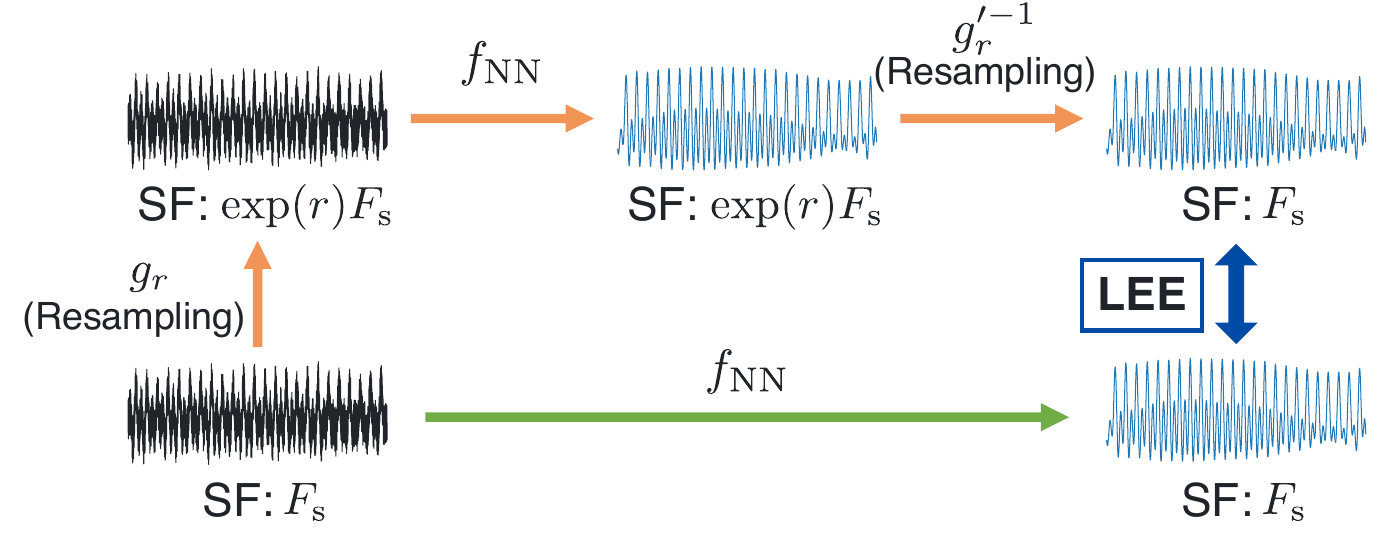}
    \caption{Schematic of LEE using resampling as input transformation.}
    \label{fig:resampling-lee}
\end{figure}
\begin{figure*}[t]
    \centering
    \subfloat[LLN-LEE]{
        \includegraphics[width=0.3\linewidth]{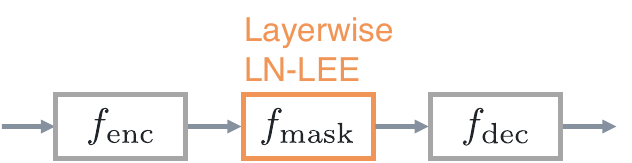}
        \vspace{\baselineskip}
        \label{fig:layerwise-lnlee}
    }
    \subfloat[$\Delta$LN-LEE]{
        \includegraphics[width=0.45\linewidth]{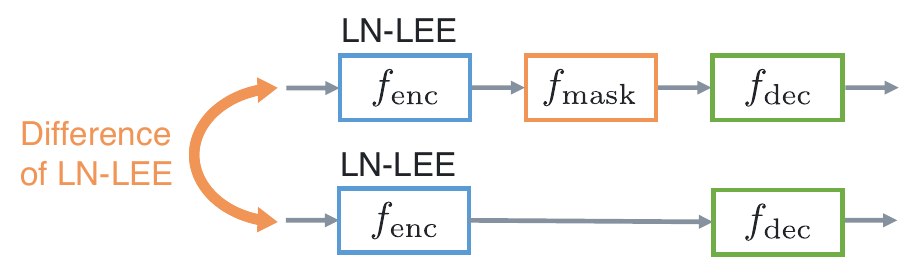}
        \label{fig:diff-lnlee}
    }
    \subfloat[Mask-LN-LEE]{
        \includegraphics[width=0.17\linewidth]{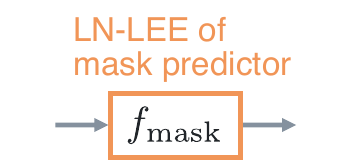}
        \label{fig:mask-lnlee}
    }
    \caption{
        Differences in DNN components evaluated in three proposed SFI evaluation metrics.
        }
        \label{fig:mask_lees}
\end{figure*}

In this section, we first apply LEE to quantify the SFI property using resampling as the input transformation.
\Cref{fig:resampling-lee} shows a schematic of LEE using resampling as the input transformation.
For brevity, we reuse $x$, $r$, and $g_r$ as the input signal, the parameter of SF change rate, and the signal resampling operation, respectively.
The length and the SF of $x$ are $N$ and $\Fs$, respectively.
We use a windowed sinc interpolation for $g_r$ to resample a signal sampled at $\Fs$ to $\exp(r)\Fs$.
To compute the LEE value, the gradient of $g_r$ with respect to $r$, denoted as $\partial g_r/\partial r$, must be computable.
Given $\Fs$, $g_r$ can be represented as a linear transformation: $g_r(x)=Sx$.
The $(m,n)$th entry of matrix $S \in \mathbb{R}^{\lceil \exp(r)N \rceil\times N}$ is given by
\begin{align}
    (S)_{m,n} &= k\left(\cfrac{1}{\Fs} \left(\cfrac{m}{\exp(r)}-n\right), \Fs\right), \label{eq:resampling}
\end{align}
where $\lceil\cdot\rceil$ denotes the ceiling function and $k(t,\Fs)$ is the following windowed sinc function:
\begin{equation}
    k(t,\Fs)=z(t)\sinc(\Fs t), \quad \sinc(t) = \dfrac{\sin(\pi t)}{\pi t}. \label{eq:sinc}
\end{equation}
Here, $t\in\mathbb{R}$ denotes the continuous time and $z(t)$ is a window function.
For $z(t)$, we choose a non-negative smooth window function and assume that $z(t) = 0$ for $t < -L/(2\Fs)$ or $t > L/(2\Fs)$, where $L$ is a positive integer.
From \cref{eq:resampling,eq:sinc}, if we choose $z(t)$ such that its gradient with respect to any $t$ is computable, we can calculate $\partial g_r/\partial r$.
That is, we can apply LEE to our problem.
For example, we can use a Hann window for $z(t)$.

To match the characteristics of DNN-based audio source separation methods, we introduce two modifications to LEE for signal resampling.
A preliminary experiment showed that the LEE value varied significantly with the scale of the DNN output.
Hence, we first normalize the LEE value by the scale of the DNN output.
Then, by taking the logarithm, similar to SDR, we define \emph{log-normalized LEE (LN-LEE)} as
\begin{equation}
    \text{LN-LEE}(\fnn) = \mathbb{E} \left[ \log_{10} \dfrac{|| \mathcal{L} \fnn(x) ||}{||\fnn(x) ||} \right].  \label{eq:ln_lee}
\end{equation}
We hereafter treat LN-LEE as a baseline evaluation metric.

\subsection{Motivation for Separate Evaluation of Mask Predictor}
LN-LEE measures the robustness of DNNs to signal resampling.
This robustness indicates that the performance of the DNN is consistent with SFs, i.e., the performance at untrained SFs does not degrade.
However, contrary to this expectation, we observed a trend that LN-LEE decreases as the performance at untrained SFs decreases, as we will show in the experimental results of \Cref{sec:result-sdr}.
This observation suggests that LN-LEE may not be suitable for evaluating the performance at untrained SFs.

To construct a metric that quantifies SFI properties and is consistent with the performance at untrained SFs, we focus on the influence of individual DNN components.
As an example network architecture, we chose the TasNet architecture as mentioned in \Cref{sec:tasnet}.
Since the encoder and decoder in this architecture correspond to time-frequency representations, the mask predictor is expected to have a greater impact on separation performance than the encoder and decoder.
Thus, we focus on the mask predictor.

Motivated by this, we propose three evaluation metrics to quantify the SFI property of the mask predictor: \emph{layerwise LN-LEE (LLN-LEE)}, \emph{difference LN-LEE ($\Delta$LN-LEE)}, and \emph{LN-LEE of the mask predictor (Mask-LN-LEE)}.
These metrics calculate the contribution of the mask predictor in different ways, and we will compare their effectiveness in \Cref{sec:exp}.
Although the proposed metrics are applicable to network architectures based on the short-time Fourier transform (STFT)~\cite{luo2023bsrnn,lu2024bsroformer,wang2023tfgridnet}, We leave the evaluation for STFT-based methods as a future work.

\subsection{LEE-based Metrics Focusing on Mask Predictor}
\noindent\textbf{LLN-LEE:} The first evaluation metric is based on layerwise LEE (see \Cref{fig:layerwise-lnlee}).
As in \Cref{sec:lee}, the Lie derivative of $\fnn$ can be decomposed as
\begin{align}
    \mathcal{L}\fnn = \mathcal{L}\fdec + \jdec \mathcal{L}\fmask + \jdec \jmask \mathcal{L}\fenc. \label{eq:layerwise_tasnet}
\end{align}
Since the contribution of the mask predictor is the second term on the right-hand side of \cref{eq:layerwise_tasnet}, the proposed metric is calculated similarly to LN-LEE:
\begin{equation}
    \text{LLN-LEE}(\fnn) = \mathbb{E} \left[ \log_{10} \dfrac{|| \jdec (\mathcal{L}\fmask)(\fenc(x)) ||}{||(\fmask\circ\fenc)(x)||} \right].
    \label{eq:layerwise_ln_lee}
\end{equation}

\smallskip\noindent\textbf{$\Delta$LN-LEE:} The second evaluation metric is based on the difference between the LN-LEE of the entire network and the network without the mask predictor (see \Cref{fig:diff-lnlee}).
In \cref{eq:layerwise_tasnet}, we treat the term of the mask predictor and the other modules separately.
From the norms of the term for the mask predictor and the other modules, we derive the following inequality:
\begin{equation}
    ||\mathcal{L}\fnn||- ||\mathcal{L}\fdec + \jdec \jmask \mathcal{L}\fenc|| \leq ||\jdec \mathcal{L}\fmask||. \label{eq:layerwise_lower}
\end{equation}
The left-hand side serves as a lower bound for $||\jdec \mathcal{L}\fmask||$ and can be used to quantify the SFI property of the mask predictor.
The left-hand side of \cref{eq:layerwise_lower} can be approximated using a network without the mask predictor, defined as $\fnomask = \fdec \circ \fenc$.
The terms $\mathcal{L}\fdec$ and $\jdec \jmask \mathcal{L}\fenc$ correspond to the contributions of the encoder and decoder, respectively.
Thus, $\mathcal{L}\fdec + \jdec \jmask \mathcal{L}\fenc$ can be approximated with the Lie derivative of $\fnomask$:
\begin{equation}
    \mathcal{L}f_\text{no\_mask} = \mathcal{L}\fdec + \jdec \mathcal{L}\fenc.
\end{equation}
With this approximation, the proposed metric is given as
\begin{equation}
    \Delta\text{LN-LEE}(\fnn) = \text{LN-LEE}(\fnn) - \text{LN-LEE}(\fnomask).
\end{equation}

Before taking the difference between $||\mathcal{L}\fnn(x)||$ and $||\mathcal{L}f_\text{no\_mask}(x)||$, we normalize the norm of each Lie derivative and take the logarithm (see \cref{eq:ln_lee}).
An alternative approach is to first compute the difference of the norms and then normalize and take the logarithm of the result.
However, preliminary experiments showed no significant difference between the two methods. 

\smallskip\noindent\textbf{Mask-LN-LEE:} The last evaluation metric is based on LN-LEE of the mask predictor (see \Cref{fig:mask-lnlee}).
We use the LN-LEE for $\fmask$ as the proposed metric:
\begin{equation}
    \text{mask-LN-LEE}(\fmask) = \mathbb{E} \left[ \log_{10} \dfrac{|| (\mathcal{L} \fmask)(\fenc(x)) ||}{||(\fmask\circ\fenc)(x)||} \right].
\end{equation}
Unlike LLN-LEE, this metric excludes the influence of $\fdec$ because $(\mathcal{L} \fmask)(\fenc(x))$ is not multiplied by $\jdec$.

\section{Experimental Analysis} \label{sec:exp}
\subsection{Experimental Setup}
We investigated the properties of the proposed evaluation metrics by applying them to music source separation methods.
We used the same experimental setup for the music source separation methods as in~\cite{saito2022sficonvtasnet}. 
We used the MUSDB18-HQ dataset~\cite{rafii2019musdbhq}, which consists of 150 tracks composed of vocals, bass, drums, and other. 
The SF of each track is 44.1~\si{\kilo\hertz}.
We used the official data split of 86, 14, and 50 tracks for training, validation, and test, respectively.
The SF for the training and validation data was set to 32 kHz.

We used SFI Conv-TasNet~\cite{saito2022sficonvtasnet} as a source separation model, which is an extension of the Conv-TasNet defined in~\cite{samuel2020metalearning} and is based on the TasNet architecture.
It uses an SFI convolutional layer and an SFI transposed convolutional layer in the encoder and decoder, respectively.
See \cite{saito2022sficonvtasnet} for details.
For latent analog filters in the SFI convolutional layer and SFI transposed convolutional layer, we utilized modulated Gaussian filter (MGF). 
Its frequency response is given by
\begin{equation}
    A(\omega) = \exp\left\{\frac{-(\omega-\mu)^2}{2\sigma^2}+\text{j}\phi\right\} + \exp\left\{\frac{-(\omega+\mu)^2}{2\sigma^2}+\text{j}\phi\right\}, \label{eq:mgf}
\end{equation}
where $\omega$ is the angular frequency, $\text{j}$ is the imaginary unit, $\mu$ is the center angular frequency, $\phi$ is the initial phase, and $\sigma$ is the parameter corresponding to the bandwidth of the filter.
We initialized $\mu$ and $\phi$ as in \cite{saito2022sficonvtasnet}.
The initialization of $\sigma$ is explained in \Cref{sec:setup-sfi}.
These parameters were trained jointly with other DNN parameters.
We utilized the frequency-domain filter design in the SFI convolutional layer and the SFI transposed convolutional layer.
The SFI Conv-TasNet was trained with a batch size of 12 for 250 epochs by RAdam~\cite{liu2020radam} wrapped in a LookAhead optimizer~\cite{zhang2019lookahead}.
We used the same data augmentations as in \cite{saito2022sficonvtasnet}.
The loss function was the negative scale-invariant source-to-noise ratio (SI-SNR)~\cite{isik2016sisnr}.

For the evaluation of the separation performance, we used the SDR obtained with the BSSEval v4 toolkit~\cite{stoter2018bsseval}.
Since the SDR is scale-dependent, we used the scaling method in~\cite{samuel2020metalearning}.
To evaluate the impact of performance degradation due to signal resampling, we used the degradation from the SDR at the trained SF to the SDR at each SF.
We call it the SDR degradation.

\subsection{Setup of Proposed Evaluation Metrics} \label{sec:setup-sfi}
We compared four evaluation metrics for the SFI property: LN-LEE for the entire network (\emph{entire LN-LEE}), LLN-LEE, $\Delta$LN-LEE, and mask-LN-LEE.
They were implemented on the basis of the official implementation\footnote{\url{https://github.com/ngruver/lie-deriv}} of LEE and calculated at the trained SF of 32~\si{\kilo\hertz}.
We extracted 5-second segments from each track in the test data. 
$z(t)$ was a Hann window and $L$ was set to 24.
To reduce the dependency on the initialization of the DNN, we used four random seeds to calculate each metric and used their average for comparison.

To evaluate how well these metrics explain the performance degradation at untrained SFs, we used the correlation coefficients $\rho$ between the metrics and the SDR degradation.
The SDRs of SFI Conv-TasNet often decrease at SFs lower than the trained SF~\cite{saito2022sficonvtasnet}.
Hence, we calculated the SDR degradations at SFs of 8 and 16~\si{\kilo\hertz}.
In preliminary experiments, we observed that the SDR degradation increases as the initial value of $\sigma$ in \cref{eq:mgf} becomes larger.
Therefore, we varied the initial values of $\sigma$ from $10\pi$ to $100\pi$ in increments of $10\pi$.

\subsection{Correlation of Proposed Metrics and Separation Performance} \label{sec:result-sdr}
\begin{figure*}[t]
    \centering
    \begin{minipage}{0.85\hsize}
        \centering
        \includegraphics[width=\linewidth]{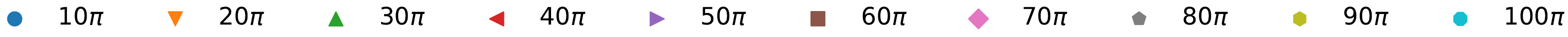}
    \end{minipage}
    \vspace{-0.8em}

    \centering
    \subfloat[Entire LN-LEE ($\rho=-0.74$)]{
        \includegraphics[width=0.23\textwidth]{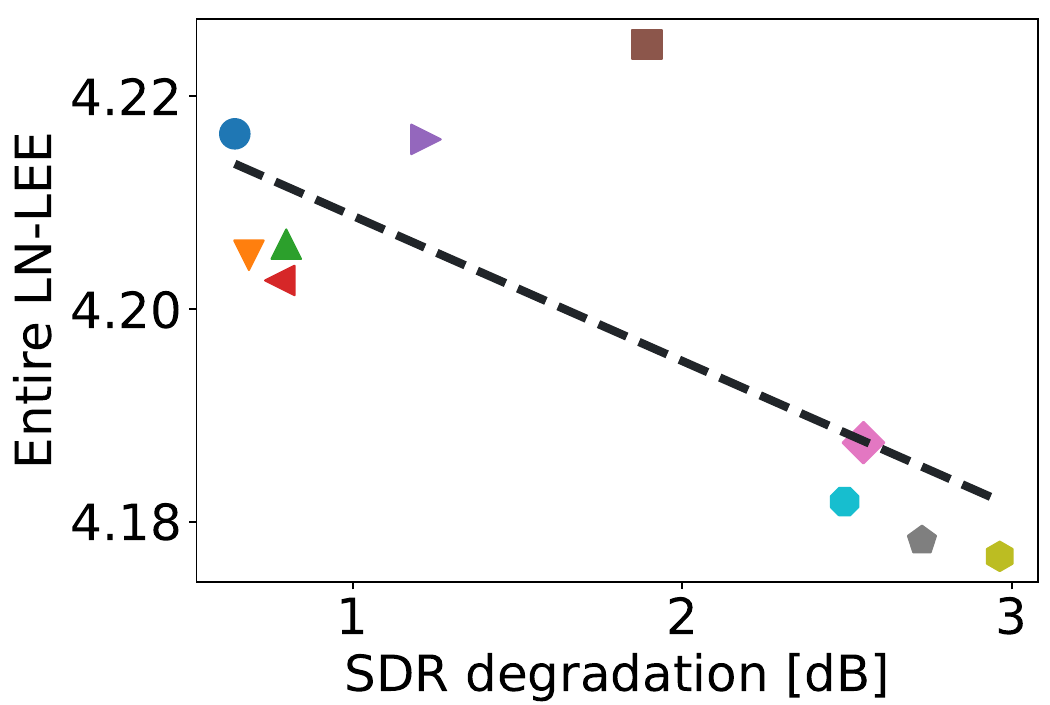}
    }
    \subfloat[LLN-LEE ($\rho=0.83$)]{
        \includegraphics[width=0.23\textwidth]{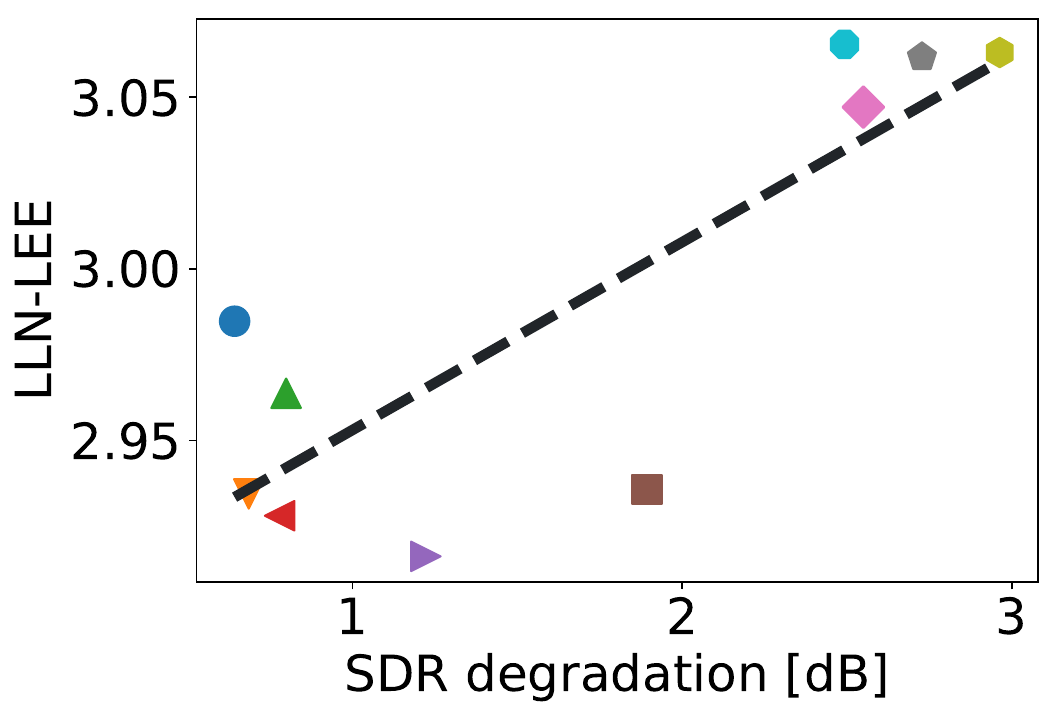}
    }
    \subfloat[$\Delta$LN-LEE ($\rho=0.91$)]{
        \includegraphics[width=0.23\textwidth]{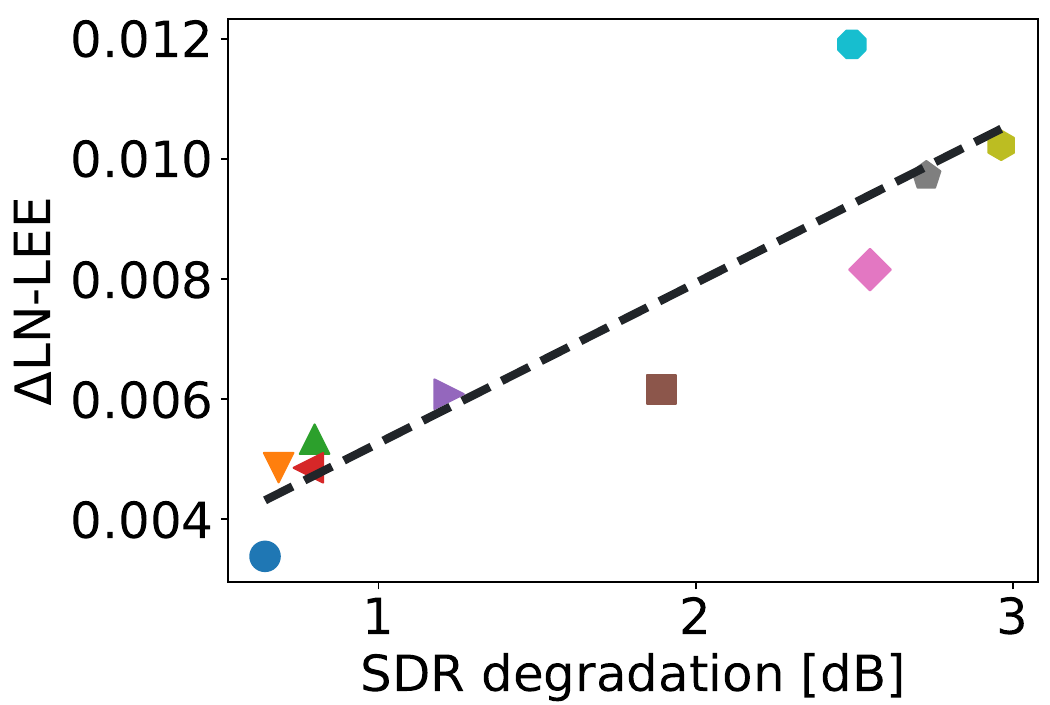}
    }
    \subfloat[Mask-LN-LEE ($\rho=0.89$)]{
        \includegraphics[width=0.23\textwidth]{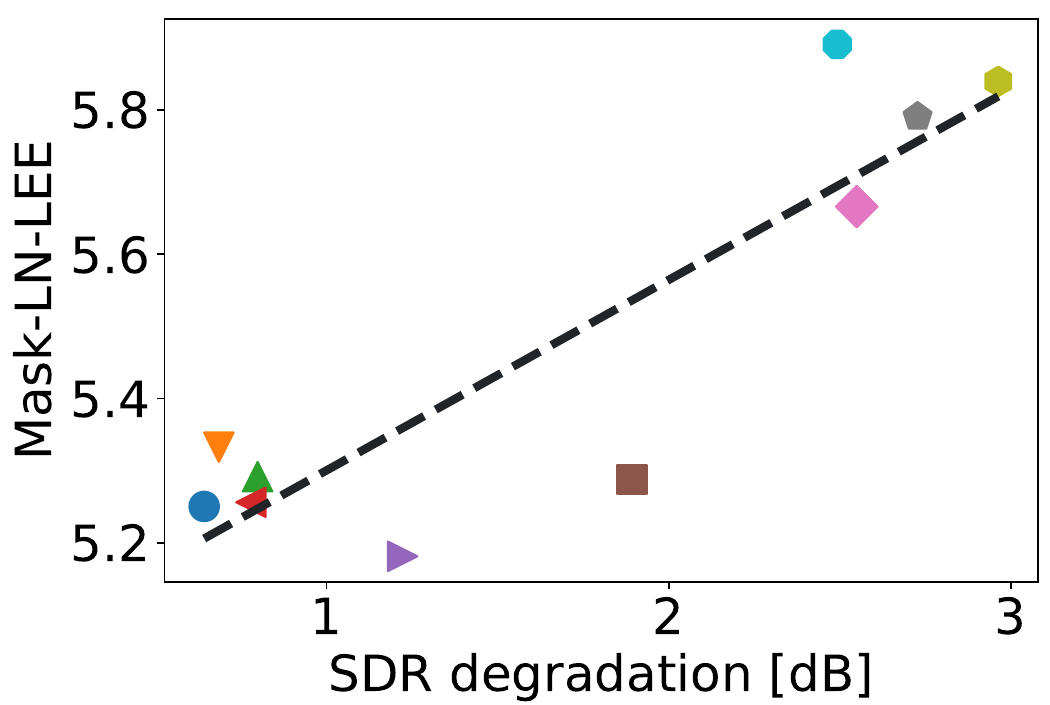}
    }
    \caption{Scatter plots of SDR degradation and each evaluation metric for SFI property at 8~\si{\kilo\hertz}. Black dashed lines denote linear regression lines.
    Each point is average of metric values calculated using models trained with four random seeds.
    }
    \label{fig:result_8khz}
\end{figure*}
\begin{figure*}[t]
    \centering
    \begin{minipage}{0.85\hsize}
        \centering
        \includegraphics[width=\linewidth]{figures/exps/legend.pdf}
    \end{minipage}
    \vspace{-0.8em}
    
    \centering
    \subfloat[Entire LN-LEE ($\rho=-0.83$)]{
        \includegraphics[width=0.23\textwidth]{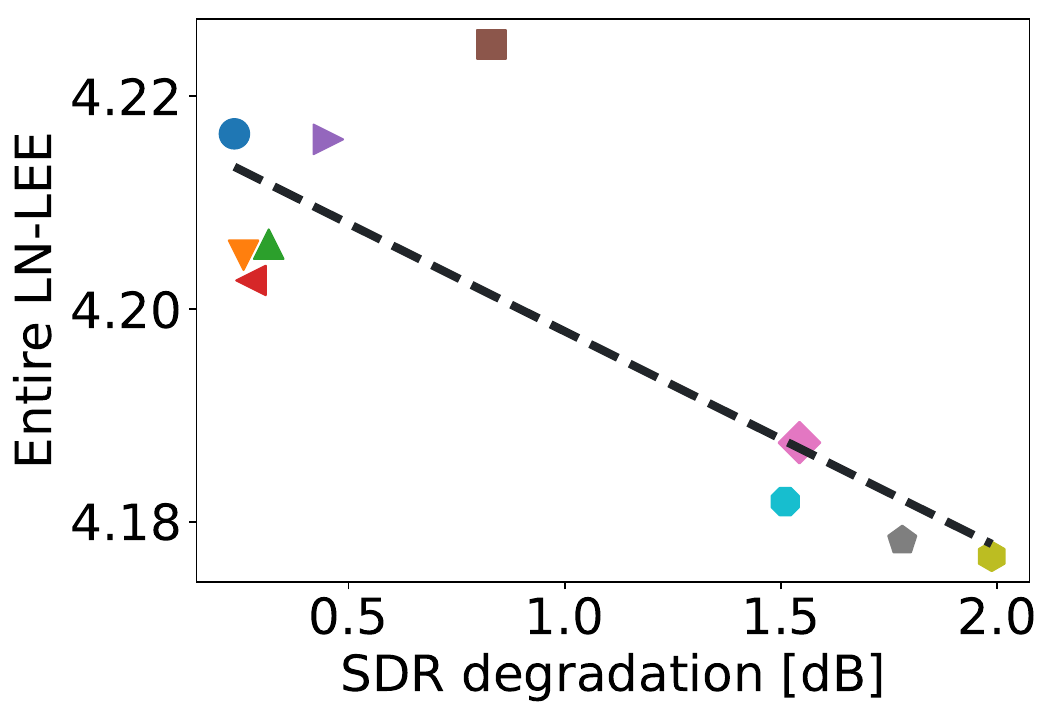}
    }
    \subfloat[LLN-LEE ($\rho=0.89$)]{
        \includegraphics[width=0.23\textwidth]{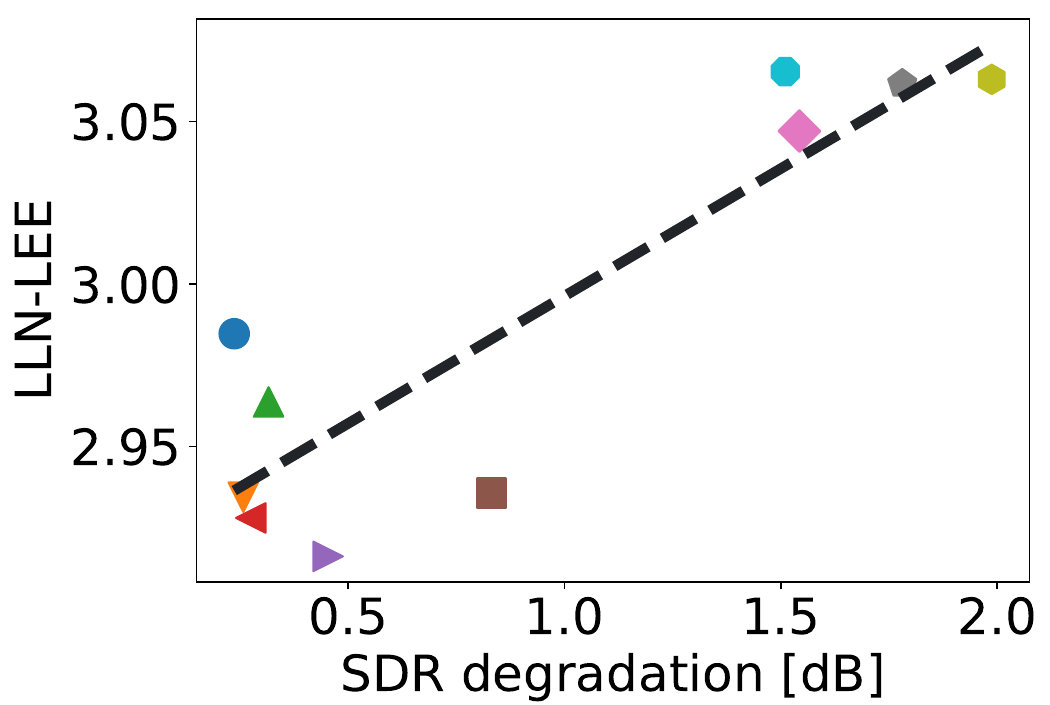}
    }
    \subfloat[$\Delta$LN-LEE ($\rho=0.91$)]{
        \includegraphics[width=0.23\textwidth]{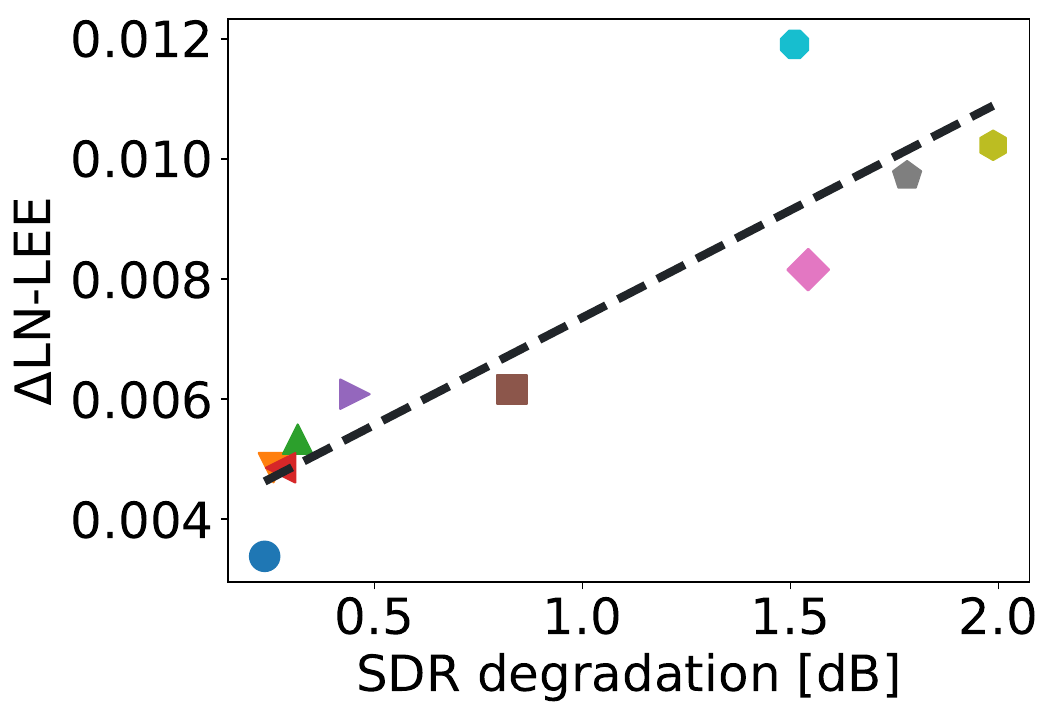}
    }
    \subfloat[Mask-LN-LEE ($\rho=0.93$)]{
        \includegraphics[width=0.23\textwidth]{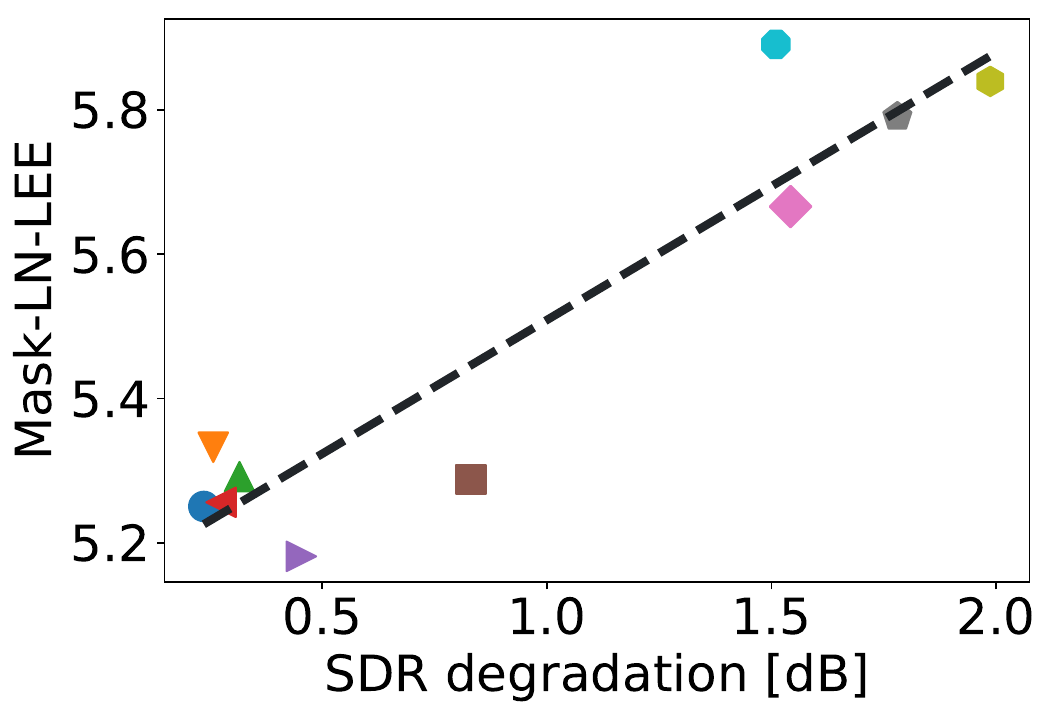}
    }
    \caption{Scatter plots of SDR degradation and each evaluation metric for SFI property at 16~\si{\kilo\hertz}. Black dashed lines denote linear regression lines.
    Each point is average of metric values calculated using models trained with four random seeds.
    }
    \label{fig:result_16khz}
\end{figure*}

\Cref{fig:result_8khz,fig:result_16khz} show scatter plots of SDR degradation and each evaluation metric at 8~\si{\kilo\hertz} and 16~\si{\kilo\hertz}, respectively.
Negative correlations were observed between the SDR degradation and entire LN-LEE for both SFs, demonstrating a contradiction with the desired trend.
For the three proposed metrics that quantify the SFI property of the mask predictor, a positive correlation with the SDR degradation was observed across all metrics and SFs.
These results demonstrate that the proposed metrics are more suitable for evaluating performance degradation at untrained SFs than entire LN-LEE.
They also highlight the importance of focusing on the mask predictor to evaluate the performance degradation at untrained SFs.

Among the three proposed metrics, mask-LN-LEE achieved the highest correlation coefficients at 16~\si{\kilo\hertz}.
With a slight difference in correlation coefficient, $\Delta$LN-LEE had similar values at 8~\si{\kilo\hertz} and 16~\si{\kilo\hertz}.
From a computational perspective, mask-LN-LEE is more efficient than $\Delta$LN-LEE because $\Delta$LN-LEE requires an additional LN-LEE computation.

Importantly, the proposed metrics were computed not at the test SFs but at the trained SF.
When we computed the proposed metric values at the test SFs, we found that the trends were similar to those in~\Cref{fig:result_8khz,fig:result_16khz}.
This result indicates that the evaluation only at the trained SF is valid across various SFs.

\section{Conclusion}
To evaluate the SFI property of DNN-based audio source separation methods, we explored evaluation metrics based on LEE using resampling as an input transformation.
Through this development, we found that directly applying LEE does not fully explain performance degradation at untrained SFs.
To resolve this issue, we proposed three metrics to quantify the SFI property in the mask predictor in the TasNet architecture.
Experiments on music source separation demonstrated that the proposed metrics are strongly correlated with the performance degradation at the untrained SFs.
They also show the importance of focusing on the mask predictor to evaluate the SFI property.
Extension of the proposed metrics to network architectures other than the TasNet architecture remains as future work.

\bibliographystyle{IEEEtran}
\bibliography{abbrev,reference}

\end{document}